\raggedbottom
\magnification=1200

\def\endsymbol{$\sqcup\mkern-12mu\sqcap$}
\def\done{\ \endsymbol\medskip}

\outer\def\beginsection#1\par{\vskip0pt plus .1\vsize\penalty-250
   \vskip0pt plus-.1\vsize\bigskip\vskip\parskip
   \message{#1}\leftline{\bf#1}\nobreak\smallskip\noindent}

\outer\def\proclaim #1. #2\par{\medbreak
   \noindent{\bof#1.\enspace}{\it#2}\par
   \ifdim\lastskip<\medskipamount \removelastskip\penalty55\medskip\fi}

\def\openR{\hbox{\rm I\kern-.2em R}}
\def\openC{\hbox{\hskip .55em}\hbox{\rm I\kern-.55em C}}
\def\smallE{{\hbox{\sevenrm E}}}
\font\titlef=cmr10 at 12pt

\centerline{\titlef   A geometric proof of the Kochen-Specker no-go theorem.}

\bigskip

\centerline{              Richard D. Gill}

\medskip

\centerline{  \it         Mathematical Institute}
\centerline{ \it          University of Utrecht}
\centerline{ \it          Budapestlaan 6}
\centerline{ \it          3584 CD Utrecht}
\centerline{ \it          Netherlands}

\medskip

\centerline{  \sevenrm           gill@math.uu.nl}

\bigskip

\centerline{             Michael S. Keane}

\medskip

\centerline{ \it         Dept.\ of Mathematics and Computer Science}
\centerline{ \it         Wesleyan University}
\centerline{ \it         Wesleyan Station Middletown}
\centerline{ \it         Connecticut 06459, USA}
\medskip

\centerline{ \sevenrm            mkeane@wesleyan.edu}

\bigskip

\centerline{Reprint of {\it J.\ Phys.\ A: Math.\ Gen.\ \bf 29} (1996), L289--L291.}

\bigskip

{

\narrower

\noindent{\bf Abstract}.  We give a short geometric proof of the Kochen-Specker
   no-go theorem for non-contextual hidden variables models.

\medskip

\noindent{\bf Key words and phrases}. Gleason's theorem, 
  noncontextual hidden variables models, quantum mechanics, 
  great circle descent.

\medskip

\noindent{\bf Note added 2 April 2003}:  I understand from 
   Jan-Aake Larsson that the construction we give here actually
   {\it contains\/} the original Kochen-Specker construction 
   as well as many others (Bell, Conway and Kochen,
   Sch\"utte, perhaps also Peres). 
   See also Larsson (2002) ``A Kochen-Specker inequality'', 
  {\it Europhysics Letters \bf 58}:799. 

}

\bigskip

\beginsection 1. Introduction.

The fundamental theorem of Kochen and Specker (1967) shows that any
hidden-variables theory for quantum measurement (on an at least
three-dimensional system) must be {\it contextual}: i.e., in a deterministic
theory, randomness is explained not just by hidden states in the quantum system
under study but also from hidden states in the measurement devices.

The theorem is usually proved by exhibiting a finite collection of
vectors in $\openC^3$ (actually, $\openR^3$ turns out to be enough)
such that it is impossible to colour each vector either red or green
subject to the following constraints: 1), within any orthogonal
triple, exactly one vector is red and the other two are green;
2), if one vector lies in a (complex) linear combination of another two and
those two are both coloured green, then it is coloured green as well.
The two constraints are connected to the so-called sum-rule and
product-rule associating values of commuting observables. For the
preparatory arguments showing why such a construction does supply
a proof of the no-go theorem for noncontextual hidden variables
models see Peres (1993) or Gill (1995a,b).

The Kochen-Specker proof is based on a construction involving $117$
vectors. Actually the heart of the construction is a special configuration
of just ten vectors which is then chained in $3$ groups of $5$ (with three
of the vectors being used twice). Ignored by most authors is an earlier construction 
of Bell (1966) again based on a special configuration of
$13$ vectors repeated a number of times. Recently Peres (1991) gave a
construction involving just $33$ vectors. In his (1993) book he also
shows a construction of Conway and Kochen involving just $31$ vectors.
This is the world record so far. Peres (1993) and Gill (1995b) also
discuss further examples due to Peres, Mermin, and others, involving
still fewer vectors, but requiring a higher-dimensional space.
A recent contribution of this kind has been made by
Cabello, Estabaranz and Garc\'{\i}a-Alcaine (1996).
Such examples do {\it illustrate\/} the Kochen-Specker theorem 
but they do not {\it prove\/} it.

Here we present a new construction similar in flavour to the Bell and
Kochen-Specker constructions, being based on a repetition of a basic
configuration. However whereas those constructions relied on some
{\it analytic\/} computations to prove their existence, our construction
relies on a {\it geometric\/} picture---in fact, exactly the same
geometric idea used by Cooke, Keane and Moran (1985) at the heart of their
elementary proof of Gleason's theorem.  The recent Peres (1991) and
Conway-Kochen (see Peres, 1993) constructions have a geometrical aspect
but are more {\it combinatoric\/}  nature. It is therefore largely a matter
of mathematical taste which proof is to be preferred. However we feel there
is some virtue in laying a connection with Gleason's theorem (which was
also the inspiration of Bell's contribution), and in having a proof which
can be `seen' from a picture without any calculation or lengthy enumeration
being necessary. Another (more complicated) geometric proof 
is given by Galindo (1976), while a more verbal proof 
using similar ideas to ours is given in the unpublished
paper Dorling (1992).

Some authors, e.g., van Fraassen (1991), use Gleason's theorem applied
to the continuum of all vectors simultaneously to (allegedly) prove the
theorem. In our opinion this cannot be built into a correct proof of the
no-go result; see Gill (1995b) for an analysis of what can go wrong.
Other authors misinterpret Bell's argument to require continuously
many vectors and hence be disqualified but this does not
do justice to Bell's argument which in our opinion is both concise and correct.

`How many vectors' are needed in a given argument seems to us a relatively
minor point. The theorem is already proved by Bell, Kochen and Specker,
and us, after the initial configuration has been shown to exist. Moreover
there are different ways of counting vectors (for instance, one might
not accept the product-rule but only use the sum rule, and thereby need
more vectors). We see no reason not to use anything at our disposal.

\beginsection 2. A geometric lemma.

Consider the one-dimensional subspaces corresponding to non-zero, real,
linear combinations of three orthogonal vectors in $\openC^k$, $k \ge 3$.
These subspaces may be represented by points on (the surface of) the
Northern hemisphere of the globe. The original triple is represented by
North pole together with two points on the equator whose longitudes differ
by $90^\circ$.

Now fix a point $\psi$ in the Northern hemisphere, not at the North pole nor
on the equator. Consider the great circle through this point which crosses
the equator at the two points differing in longitude by $\pm 90^\circ$
from $\psi$. Choose one of these equatorial points and
call it $\psi^\smallE$.
Call the point on the Northern hemisphere orthogonal to the great circle
$\psi^\perp$. Its longitude is that of $\psi$ plus $180^\circ$ and its latitude
is $90^\circ$ minus that of $\psi$. The triple $\psi$, $\psi^\smallE$,
$\psi^\perp$ are orthogonal.

The great circle we just defined has $\psi$ as its most Northerly point. We
call it {\it the great circle descent from $\psi$}.

Starting from a point $\psi=\psi_0$ go down its descent circle some way
to a new point $\psi_1$. Now consider the new great circle descent from
$\psi_1$. Go down some way to a new point $\psi_2$, and so on. After $n$
steps arrive at $\psi_n$. Obviously $\psi_n$ is more Southerly than $\psi_0$.
Cooke, Keane and Moran's geometric lemma states that one can reach {\it any\/}
more Southerly point than $\psi_0$ by a finite sequence of great circle
descents. For instance, one can fly from Amsterdam to Tokyo by a finite
sequence of great circle descents.

The lemma is proved by projecting the Northern hemisphere {\it from\/} the
centre of the earth {\it onto\/} the horizontal plane tangent to the
earth at the North pole. Lines of constant latitude project onto concentric
circles, a great circle descent projects onto a straight line tangent to
the circle of constant latitude at its summit.

\beginsection 3. Proof of the theorem.

Start with an orthogonal triple. Colour one point red and the other
two green. Let the red point be the North pole and the other two green
points be on the equator. Any further points selected on the equator get
coloured green by the product rule. Take a point $\psi$ at latitute
$60^\circ$ above the equator. Together with $\psi^\perp$ and
$\psi^\smallE$ we have a new orthogonal triple. Since $\psi^\smallE$
gets coloured green, if $\psi$ is coloured green then $\psi^\perp$ is coloured
red. Note that $\psi^\perp$ lies at $30^\circ$ above the equator, more
Southerly than $\psi$.

Suppose $\psi$ is coloured green. Since any point on its great circle descent
is a linear combination of $\psi$ and $\psi^\smallE$, it is also coloured
green. Repeating this argument, any point which can be reached by a finite
number of great circle descents from $\psi$ is also coloured green. But this
applies to $\psi^\perp$, a contradiction.

Therefore $\psi$ is coloured red just like the North pole. So we have shown
that any point within $30^\circ$ of a red point is also coloured red.
Now go in three steps of $30^\circ$ from the North pole down to the equator,
then in three steps of $30^\circ$ along the equator, then in three steps
of $30^\circ$ back up to the North pole. One of the three `corners' of this
circuit has to be coloured red, hence they all are,
a contradiction. \done

\beginsection References.

\def\ref{\hang\noindent}

\ref J.S. Bell (1966), On the problem of hidden variables in quantum mechanics, {\it
Rev.\ Mod.\ Phys.\ \bf 38}, 447--452.

\ref A. Cabello, J.M. Estebaranz and G. Garc\'{\i}a-Alcaine (1996),
Bell-Kochen-Specker theorem: a proof with 18 vectors, 
{\it J. Phys.\ Lett.\ A \bf 212}, 183--187.

\ref R. Cooke, M. Keane and W. Moran (1985), An elementary proof of Gleason's theorem,
{\it Math.\ Proc.\ Camb.\ Phil.\ Soc.\ \bf 98}, 117--128.

\ref J. Dorling (1992), {\it A simple logician's guide to the quantum
puzzle and to quantum logic's putative solution\/}, preprint,
Univ.\ Amsterdam.

\ref B.C. van Fraassen (1991), {\it Quantum Mechanics: An Empiricist View}, Oxford
University Press.

\ref A. Galindo (1976), Another proof of the Kochen-Specker theorem,
{\it Algunas cuestiones de fisica teorica\/ 1.975} (In memoriam
Angel Esteve), 3--9, GIFT, Zaragoza.

\ref R.D. Gill (1995a), {\it Discrete Quantum Systems}, Preprint, Dept.\ Math., Univ.\
Utrecht.

\ref R.D. Gill (1995b), {\it Notes on Hidden Variables}, Preprint, Dept.\ Math., Univ.\
Utrecht.

\ref S. Kochen and E.P. Specker (1967), The problem of hidden variables in quantum
mechanics, {\it J. Math.\ Mech.\ \bf 17} 59--87.

\ref A. Peres (1991), Two simple proofs of the Kochen-Specker theorem, {\it J. Phys.\
A: Math.\ Gen.\ \bf 24} (Letter to the editor), L175--L178.

\ref A. Peres (1993), {\it Quantum Theory: Concepts and Methods}, Kluwer, Dordrecht.

\bye